\documentclass{iau} 
\usepackage{graphicx}	% Including figure files
\usepackage{amsmath}	% Advanced maths commands
\usepackage{amssymb}	% Extra maths symbols
\usepackage{upgreek}
\usepackage{booktabs}
\usepackage{mathrsfs}
\usepackage{calrsfs}
\usepackage{hyperref}
\hypersetup{colorlinks=false}
\usepackage{pdflscape}
\usepackage{rotating}
\usepackage{longtable}
\usepackage{multicol}

\title[Using low-frequency pulsar observations to study the Galactic magnetic field] %% give here short title %%
{Using low-frequency pulsar observations to study the 3-D structure of the Galactic magnetic field}

\author[C. Sobey et al.]   %% give here short author list %%
{C. Sobey$^{1,2}$
 \and the LOFAR and MWA collaborations}

\affiliation{$^1$International Centre for Radio Astronomy Research - Curtin University, GPO Box U1987, Perth, WA 6845, Australia\\[\affilskip]
$^2$ CSIRO Astronomy and Space Science, 26 Dick Perry Avenue, Kensington, WA 6151, Australia \\email: {\tt c.sobey@curtin.edu.au}}

\pubyear{2018}
\volume{333}  %% insert here IAU Symposium No.
\jname{Peering towards Cosmic Dawn}
\editors{Vibor Jeli\'c \& Thijs van der Hulst, eds.}
\begin{document}

\maketitle

\begin{abstract}
The Galactic magnetic field (GMF) plays a role in many astrophysical processes and is a significant foreground to cosmological signals, such as the Epoch of Reionization (EoR), but is not yet well understood. 
Dispersion and Faraday rotation measurements (DMs and RMs, respectively) towards a large number of pulsars provide an efficient method to probe the three-dimensional structure of the GMF. 
Low-frequency polarisation observations with large fractional bandwidth can be used to measure precise DMs and RMs. This is demonstrated by a catalogue of RMs (corrected for ionospheric Faraday rotation) from the Low Frequency Array (LOFAR), with a growing complementary catalogue in the southern hemisphere from the Murchison Widefield Array (MWA). These data further our knowledge of the three-dimensional GMF, particularly towards the Galactic halo. 
Recently constructed or upgraded pathfinder and precursor telescopes, such as LOFAR and the MWA, have reinvigorated low-frequency science and represent progress towards the construction of the Square Kilometre Array (SKA), which will make significant advancements in studies of astrophysical magnetic fields in the future. A key science driver for the SKA-Low is to study the EoR, for which pulsar and polarisation data can provide valuable insights in terms of Galactic foreground conditions. 
\keywords{ISM: magnetic fields, Galaxy: halo, pulsars: general, polarization }
\end{abstract}

\firstsection % if your document starts with a section,
              % 6 page limit
\section{Introduction}

%\cite[Anders \& Zinner (1993)]{AndersZinner93} and 
%\cite[Ott (1993)]{Ott93}.
% 

Magnetic fields are ubiquitous throughout the Universe. They play a role in numerous astrophysical processes across a range of physical scales and field strengths: from magnetar surfaces with up to petagauss field strengths (e.g. Tiengo et al. 2013); to intergalactic magnetic fields with less than 30 femtogauss field strengths (e.g. Essey et al. 2011). The magnetospheres of pulsars with $\sim$teragauss field strengths are involved in generating radiation (e.g. Goldreich \& Julian 1969) across the electromagnetic spectrum: from the lowest radio frequencies observable from the Earth's surface ($\sim$10MHz; e.g. Hassall et al. 2012) up to high-energy gamma rays ($>$100GeV; e.g. Aliu et al. 2011). The geomagnetic field ($\sim$1\,G) deflects charged particles in the solar wind and cosmic rays that would otherwise erode the Earth's atmosphere.  Galactic magnetic fields with $\sim\upmu$G field strengths pervade the diffuse ISM and influence many processes in the Galactic ecosystem, e.g., cosmic ray acceleration and confinement over large scales (e.g. Aharonian et al. 2012, and references therein) and molecular-cloud and star formation over smaller scales (e.g. Crutcher 2012, and references therein).

Despite decades of study, magnetic fields in the Universe are still not well understood. For example, the mechanism by which the pulsar magnetosphere generates and varies its emission remains obscure, and the 3-D structure of the GMF is still under debate. 
However, work towards better understanding these magnetic fields is ongoing and has been reinvigorated by the recent construction of next-generation low-frequency radio telescopes, including LOFAR (van Haarlem et al. 2013), the Long Wavelength Array (LWA; Taylor et al. 2012), the MWA (Tingay et al. 2013), and their associated supercomputing facilities, as well as recent upgrades to existing telescopes, e.g., GMRT (Gupta et al. 2017).

Here, we describe the GMF and methods used to investigate it (Section 2). We outline recent and ongoing work using the LOFAR and MWA telescopes to measure Faraday rotation towards pulsars to study the GMF in 3-D (Section 3).  Finally, we summarise and look forward to the era of the Square Kilometre Array (SKA) in the future (Section 4).

\section{The Galactic magnetic field}

%The galactic magnetic field

Our Galaxy provides a unique opportunity for high-resolution measurements of a galactic magnetic field, for comparison with nearby galaxies (e.g. Nota \& Katgert 2010).  The GMF affects the heat, momentum and cosmic ray energy transfer within the Galaxy, and presents a significant foreground for a wide variety of extragalactic studies (e.g. Haverkorn et al. 2015; Ghosh et al. 2017, and references therein).  Therefore, measuring and understanding the 3-D GMF is important for understanding our Galactic ecosystem and also to discern cosmological signals in the data.

The GMF is often categorised into a large-scale ordered component coherent over kiloparsec length scales, and a small-scale random component, caused by turbulence, (supernova) explosions and other localised phenomena (e.g. Jaffe et al. 2009).  Our current picture of the large-scale GMF is that the field strength is of the order of a few microgauss at the solar radius, increasing towards the Galactic centre (e.g. Han et al. 2006; Eatough et al. 2013). Recent analyses often favour an axisymmetric spiral structure with an overall clockwise direction and one field reversal near the Sun that seems to somewhat follow the spiral-arm structure (e.g. Van Eck et al. 2011; Jansson \& Farrar 2012). However, these reversals have not been observed in other galaxies, and the number and locations of the field reversals are still under debate (e.g. Han et al. 2006). The 3-D large-scale GMF in the halo is even less well understood, with several proposed geometries (see Haverkorn et al. 2015, and references therein), including a north--south asymmetry across the Galactic disk (e.g. Mao et al. 2012). 

We are able to study magnetic fields in astrophysical plasmas by observing their effects on radiation. For instance, the magnetic field in sunspots was first identified through Zeeman splitting in Solar spectra (Hale 1908). The GMF was first measured using polarisation of starlight (Hall 1949; Hiltner 1949). Other observables that can also be used to infer the magnetic field strength and/or direction in the GMF include the (polarised) synchrotron radiation of the Galactic non-thermal radiation (e.g. Sun \& Reich 2010, and references therein); polarised thermal dust emission (e.g. Adam et al. 2016, and references therein); and the Faraday rotation effect (e.g. Cooper \& Price 1962; Van Eck et al. 2011).  These observables are all complementary, often tracing different phases and components of the ISM and, using ancillary data, mostly delivering 2-D information about each line-of-sight (e.g. Haverkorn 2015). Here, we focus on measuring Faraday rotation towards pulsars.

Pulsars are highly magnetised and rapidly rotating neutron stars, first discovered to emit at radio wavelengths by Jocelyn Bell Burnell using a fixed dipole array that observed the sky at 81.5 MHz (Hewish et al. 1968, Pilkington et al. 1968). Since then, 2620 more pulsars have been discovered (see the pulsar catalogue\footnote{pulsar catalogue (v. 1.57) retrieved from \href{http://www.atnf.csiro.au/people/pulsar/psrcat/}{http://www.atnf.csiro.au/people/pulsar/psrcat/}}; Manchester et al. 2005).  The pulsed and often polarised nature of pulsar emission, plus their distribution throughout the Galaxy (with distances estimated using a Galactic electron density model, e.g. Yao et al. 2017, or measured via parallax), makes them excellent probes of the ionised interstellar medium. Determining dispersion and Faraday rotation measures (DMs and RMs, respectively) from their emission provides an efficient method for ascertaining the average magnetic field strength and direction (parallel to the line of sight) in the intervening media. A large set of pulsars with DM and RM measurements allows us to probe the 3-D structure of the GMF (e.g. Manchester 1972; Manchester 1974; Rand \& Lyne 1994; Han et al. 1999, 2006; Noutsos et al. 2008). Currently, 734 (28\%) known pulsars have published RMs, the majority of which have been measured at L-band radio frequencies and are located towards the Galactic disc (see the pulsar catalogue; Manchester et al. 2005). 
In addition to pulsars that provide 3-D data about the GMF, our picture of the Faraday sky also includes 41,632 RM measurements from extragalactic sources (Oppermann et al. 2015) that provide data for lines-of-sight through the entire Galaxy for comparison (although with more measurements in the northern sky; e.g. Taylor et al. 2009).

\section{Low-frequency pulsar observations}

Low-frequency polarisation observations facilitate precise measurements of DMs and RMs, due to the wavelength-squared dependence on the pulse arrival delay and polarisation angle rotation, respectively. 
We capitalise on the RM precision potential of these low-frequency data by employing the powerful method of RM-synthesis (Burn 1966; Brentjens \& de Bruyn 2005).
This method is now increasingly utilised to extract RMs from polarisation data (e.g. Lenc et al. 2017; van Eck et al. 2017) because the entire bandwidth can be used simultaneously to coherently add the complex polarisation vectors from the Stokes $Q,U$ parameters and obtain Faraday spectra. The precision available from the low-frequency data can be characterised by the FWHM of the RM spread function in Faraday space ($\delta\phi$; analogous to a point spread function in sky coordinates). For data with a frequency range of $\approx$110--190\,MHz, $\delta\phi\approx$1\,rad m$^{\rm{-2}}$, compared to $\delta\phi\approx$300\,rad m$^{\rm{-2}}$ at 1.3--1.5\,GHz.

The ionosphere, as well as the ISM, is a magneto-ionic medium and introduces additional time and position-dependent Faraday rotation.    We are currently correcting the total RM measured for the ionospheric contribution using the publicly available \textsc{ionFR} code (Sotomayor-Beltran et al. 2013).  This code models the ionospheric RM for a specified line-of-sight using publicly-available total electron content maps and the International Geomagnetic Reference Field.  Although the output of the code was verified using LOFAR RM measurements towards several pulsars (see Sotomayor-Beltran et al. 2013), this is generally the most substantial contribution to the cumulative uncertainties ($\approx$0.1\,rad m$^{\rm{-2}}$) at low frequencies. 

LOFAR's high-time and -frequency-resolution tied-array observations (Stappers et al. 2011) produce high-quality polarisation profiles of pulsars at low frequencies ($\le$200\,MHz; Noutsos et al. 2015). Full polarisation data for a large set of pulsars are available, including the high-band antenna (HBA) `slow' pulsar census (158 detected; Bilous et al. 2016) and the millisecond pulsar (MSP) census (48 detected; Kondratiev et al. 2016). LOFAR has also discovered over sixty new pulsars (in the LOFAR Tied-Array All-Sky Survey; Sanidas et al. in prep.) and continues to time these new discoveries and a subset of known slow pulsars and MSPs, providing data for an increasing number of lines-of-sight through the Galaxy.  These data are usually taken at a centre frequency of 150 MHz, with 80 MHz bandwidth, and with at least 20-minute integrations, yielding a predicted noiseless $\delta\phi\approx$0.8\,rad m$^{\rm{-2}}$. The data analysed to date have provided a catalogue of precise, and ionosphere-corrected, low-frequency RMs towards $\approx$200 pulsars, mostly toward the Galactic halo and in the range --5 to +5\,$\upmu$G (Sobey et al. in prep.). Approximately 90 of these do not have a previously published RMs, and the remaining measurements are $\approx$30$\times$ more precise compared to published values (generally from data above 1\,GHz). 

Our Galaxy cannot be observed in its entirety from one point on the Earth and, therefore, the MWA is being utilised to measure RMs in the southern sky. 
The MWA routinely observes pulsars (e.g. Bhat et al. 2016; McSweeney et al. 2017; Meyers et al. ApJ in press) in high-time and -frequency resolution mode (Tremblay et al. 2015). An initial total intensity pulsar census has been carried out (Xue et al. PASA in press), and the polarisation profiles are being verified (Xue et al. in prep.). Initial observations have provided RM measurements (Sobey et al. in prep.) that also supply excellent resolution in Faraday space: $\delta\phi\approx$0.4--6.2 rad m$^{\rm -2}$, depending on the centre frequency (89--216\,MHz; e.g. Lenc et al. 2017).  Therefore, the MWA and LOFAR provide complementary resolution in Faraday space and together facilitate an all-sky picture of low-frequency RMs.

LOFAR and the MWA can each be used as an interferometer to image the sky or as a tied-array (i.e. coherently summing signals from each station/tile), and both modes are useful for this work. RMs towards pulsars have also been measured in low-frequency imaging data because of the steep spectral indices and high polarisation degrees of pulsar emission (e.g. Lenc et al. 2017). Low-frequency images will be useful for detecting pulsars at low frequencies that cannot be detected in the tied-array data due to large DMs and scattering measures (e.g. Kondratiev et al. 2016) while retaining the RM measurement precision.  Furthermore, LOFAR images of the EoR field centred on 3C196 field discovered a pulsar candidate that was later followed-up and confirmed using the high-time and -frequency resolution mode (Jeli\'c et al. 2015). In turn, this pulsar provides further information within the EoR field, including the electron density, and the electron-density weighted average magnetic field strength and direction parallel to the line-of-sight.

LOFAR and the MWA provide precise measurements of the ISM properties towards pulsars in the Galactic halo and nearby pulsars in the Galactic disc, which further our knowledge of the 3-D GMF structure and can be used to constrain current models suggested for the Galactic halo. Accurate reconstruction of the GMF is increasingly hampered by the lack of independent distance measurements, e.g., parallaxes and globular cluster associations, and the uncertain distance estimates. However, we are entering an era where precision DM and RM measurements are becoming routine, allowing us to monitor smaller temporal and spatial variations (to study, e.g., small-scale turbulent ISM structures, globular clusters, the heliosphere (Howard et al. 2016), and the ionosphere) even more pertinent to EoR fields.

\section{Summary and future work}

We described how magnetic fields play a role in many astrophysical processes across many orders of magnitude in scale and field strength.
This includes the Galactic magnetic field (GMF), which is a significant foreground to the EoR signal, and for which pulsar observations can be used to probe its 3-D structure. Low-frequency observations using LOFAR and the MWA facilitate all-sky coverage and precise measurements towards reconstructing the GMF, particularly towards the Galactic halo. It is essential to correct the observed RM for the ionospheric Faraday rotation in order to achieve accurate measurements for the ISM alone (e.g. Sotomayor-Beltran et al. 2013). We are now entering a monitoring era, where precise DMs/RMs can be used to investigate smaller-scale and turbulent structures in the ISM.  Ongoing pulsar surveys in the time domain, e.g. the LOFAR Tied-Array All-Sky Survey (Sanidas et al. in prep.), are discovering more pulsars and, therefore, more lines-of-sight with which we can probe and reconstruct the GMF. New techniques are also being developed to detect pulsar candidates in the imaging domain, e.g. Dai et al. 2017.

The SKA's capabilities for observing and discovering pulsars will revolutionise radio astronomy, including studies of astrophysical magnetic fields, e.g., the 3-D GMF structure on large and small scales (Han et al. 2015). 
A good understanding of the polarisation performance of the antennas and accurate ionospheric monitoring will be invaluable for many science cases, including polarisation and EoR studies. 
The data obtained from pulsar observations will be complementary to the `RM grid' of extragalactic point sources and diffuse Galactic synchrotron images, towards a more comprehensive picture of the GMF and understanding our Galaxy as a (3-D) foreground to extragalactic and cosmological signals (Haverkorn et al. 2015; Haverkorn, these proceedings).

\begin{acknowledgements}

Thank you to all of our collaborators. 
%Thank you to the LOC for organising a fantastic conference, and to the SOC for inviting me to give a talk.
Results discussed are based on observations using the International LOFAR Telescope and the Murchison Widefield Array. 
LOFAR (van Haarlem et al. 2013) is the Low Frequency Array designed and constructed by ASTRON. 
It has facilities in several countries, that are owned by various parties (each with their own funding sources), and that are collectively operated by the ILT foundation under a joint scientific policy.
This scientific work makes use of the Murchison Radioastronomy
Observatory, operated by CSIRO. We acknowledge the
Wajarri Yamatji people as the traditional owners of the Observatory
site. 
Support for the operation of the MWA is provided by the Australian Government (NCRIS), under a contract to Curtin University administered by Astronomy Australia Limited. 
We acknowledge the Pawsey Supercomputing Centre which is supported by the Western Australian and Australian Governments. 
We acknowledge the International Centre for Radio Astronomy Research (ICRAR), a Joint Venture of Curtin University and The University of Western Australia, funded by the Western Australian State government.

\end{acknowledgements}

\end{document}